\input{aipcheck}
  
\documentclass[final]{aipproc}

\layoutstyle{6x9}

\begin{document}

\title{Light radioactive nuclei capture reactions with
phenomenological potential models.}


\author{V. Guimar\~aes}
{address={Instituto de F\'{\i}sica, Universidade de S\~ao Paulo, 
P.O. Box 66318, 05389-970 S\~ao Paulo, SP, Brazil},
email=valdirg@dfn.if.usp.br}

\author{C. A Bertulani}
{address={Texas A$\&$M University-Commerce, Commerce, Texas 75429, USA.}}

\begin{abstract}
Light radioactive nuclei play an important role in many astrophysical
environments.  
Due to very low cross sections of some neutron and proton capture 
reactions by these radioactive nuclei at energies of astrophysical interest, 
direct laboratory  measurements are very difficult. For radioactive nuclei 
such as $^8$Li  and $^8$B, the direct measurement of neutron capture 
reactions is impossible.  Indirect methods have been applied to overcome
these difficulties.
In  this work we will report on the results and discussion of phenomenological 
potential models used to determine some proton and neutron capture reactions.
As a test we show the results for the 
$^{16}$O(p,$\gamma$)$^{17}$F$_{gs}$(5/2$^+$) and 
$^{16}$O(p,$\gamma$)$^{17}$F$_{ex}$(1/2$^+$) capture reactions.
We also computed the nucleosynthesis cross sections
 for the $^7$Li(n,$\gamma$)$^8$Li$_{gs}$,
 $^8$Li(n,$\gamma$)$^9$Li$_{gs}$ and  $^8$B(p,$\gamma$)$^9$C$_{gs}$
capture reactions. 
\end{abstract}

\maketitle

\section{Introduction}
\par\hfill

In the investigation of many astrophysical entities such as
primordial universe, main path stellar evolution, novae, super-novae
explosion, X-ray bursts etc. , the important
input parameter in the models is the cross sections of the capture
reactions as a function  of energy. However, since in many of these 
environments the temperature is not very high, the cross sections have to be 
obtained  at the ``{\it Gamow Peak}'' energies which are very low, in the range 
of few  tens to at most hundreds of keVs. 
Whereas some of the cross sections originate from laboratory 
measurements,  the majority are based on extrapolations from the 
higher to lower energies or are obtained by pure theoretical 
models with, sometimes, no firm experimental basis. Some reactions
are practically impossible to be directly measured. 
For instance, for light radioactive nuclei nucleosynthesis, the 
neutron capture reaction is impossible because the combination of target+beam 
is not possible. A typical case is the  $^8$Li(n,$\gamma$)$^9$Li capture 
reaction where direct
measurement is not possible because no $^8$Li or neutron target exist.
For proton capture reactions involving radioactive nuclei such as 
$^8$B or $^7$Be, the direct measurement is very difficult due to the 
low cross sections and limited beam intensities of these elements.
For such cases, indirect methods to determine the cross sections were
developed and are quite usually adopted.  Extrapolation from higher energy 
data to lower regime is also not straight forward.   
Careful and accurate account of physically relevant information  has 
to be considered in the description of the reaction before the extrapolation 
is performed, where to describe  capture reactions 
not only information on the structure of the nuclei,
but also a clear understanding of the reaction mechanism is
required. 

Many indirect methods have been developed to overcame the difficulties
to obtain low energies cross sections. Among these methods
we have the  Coulomb dissociation method,  which corresponds 
to the inverse temporal reaction of the capture \cite{baur94}, 
the trojan horse method \cite{spi03},
the reduced-width or ANC (Asymptotic 
Normalization Coefficient) method \cite{gag02} and potential model, 
where the latter two
use transfer reactions as a way to get information on the 
non-resonant part of the capture reaction process and will be 
discussed in more detail in the next section. 
These indirect methods are very suitable to be used in association with 
low-energy radioactive nuclear beams.

\section{light radioactive nuclei and nucleosynthesis.}
\par\hfill

To overcame the A=8 gap and synthetize heavier elements the key 
nuclei are the light radioactive elements $^8$Li(178 ms) and $^8$B(770 ms).
These elements are important for instance in the 
inhomogeneous big-bang nucleosynthesis (IBBN) \cite{mal89}.
Because of the assumption of homogeneous baryon density distribution and 
also the  instability of $^8$Be, the nuclear reaction flows stop at A=8 
in SBBN (Standard Big Bang Nucleosynthesis).
In IBBN the baryon density distribution is assumed to be inhomogeneous due 
to several cosmological processes before an onset of primordial 
nucleosynthesis. A difference in diffusion between protons and neutrons 
makes high density zones proton-rich and low density zones neutron-rich. 
In such an environment, heavy elements can be produced via the formation 
proton or neutron rich unstable nuclei. 
Also it is believed that in supermassive stars with low metalicity, 
the 3$\alpha$ capture process cannot
form enough $^{12}$C to initiate the rapid hydrogen burning during the 
explosion.
However, in such high temperature and densities the hot $pp$-chain
my bypass the slower 3$\alpha$ capture \cite{wie89} and in this 
case $^8$B plays an important role.

The following nuclear reaction 
chains were found to be very important in nucleosynthesis processes
discussed above to jump
the A=8 gap and synthetise heavier elements:
\par\hfill

\noindent {\bf chain (1):}  
$^7$Li(n,$\gamma$)$^8$Li($\alpha$,n)$^{11}$B(n,$\gamma$)$^{12}$B(e$^-\nu$)$^{12}$C.

\noindent {\bf chain (2):}  
$^7$Li(n,$\gamma$)$^8$Li(n,$\gamma$)$^9$Li(e$^-\nu$)$^{9}$Be(n,$\gamma$)$^{10}$Be(e$^-\nu$)$^{10}$B(n,$\gamma$)$^{11}$B(n,$\gamma$)$^{12}$B(e$^-\nu$)$^{12}$C.

\noindent {\bf chain (3):}  
$^7$Be(p,$\gamma$)$^8$B($\alpha$,p)$^{11}$C(e$^+\nu$)$^{11}$B(p,$\gamma$)$^{12}$C.

\noindent {\bf chain (4):}  
$^7$Be(p,$\gamma$)$^8$B($\alpha$,p)$^{11}$C(p,$\gamma$)$^{12}$N($\beta$)$^{12}$C.

\noindent {\bf chain (5):}  
$^7$Be(p,$\gamma$)$^8$B(p,$\gamma$)$^{9}$C($\alpha$,p)$^{12}$N($\beta$)$^{12}$C.
\par\hfill

The first two reaction chains play the central role in the 
production of neutron-rich nuclei, and the chains (3), (4) and (5) of 
proton-rich nuclei.
The $^8$Li(n,$\gamma$)$^{9}$Li is also an important reaction 
in the early stage of the ignition of supernovae, where two neutrons
capture reactions,$^9$Li(2n,$\gamma$)$^{11}$Li and 
 $^6$He(2n,$\gamma$)$^{8}$He($\beta$)$^8$Li, may become important. 
The  $^{7}$Be(p,$\gamma$)$^{8}$B is related to the solar neutrino
problem. Precise predictions of the production rate of $^8$B solar neutrinos 
are  important for testing solar models, and for limiting the allowed neutrino 
mixing parameters. This, however, it the most uncertain reaction leading to 
$^8$B formation in the Sun \cite{Junghans03}.
The $^8$B(p,$\gamma$)$^{9}$C, capture reaction 
is also important in the novae environment where  temperatures are
several times larger than $10^8$ K, corresponding to
Gamow window energies around $E=50 - 300$ keV \cite{wie89,ful86}.


\section{$S$-factor at low energies}
\par\hfill

Cross sections for many of the neutron and proton capture reactions by 
light  radioactive  nuclei are still 
poorly known or remain unmeasured at the required low energy region.
 For systems which cannot be measured directly at low energies
some procedures or methods are adopted to obtain cross sections
in this region. The most obvious way to obtain the low energy
cross section is the extrapolation to low energy through polynomial 
parametrization of the  available high energy region data.
Actually these extrapolations are performed to the $S$-factor where 
the strong energy dependence  of charged  particle capture reactions 
due to the coulomb barrier penetration are removed. 
Thus, the cross sections are conventionally expressed in terms of the 
$S$-factor which is defined as:

\begin{equation}
S(E) =E \sigma(E) \exp[ 2\pi\eta(E)] \hspace{1cm}
{\rm with} \hspace{1cm} \eta(E)=Z_{a}Z_{b}e^{2}/\hbar v,
\end{equation}

\noindent where $v$ is the initial relative velocity between the nucleus and
proton.

However, the $S$-factor does not remove completely the energy dependence,
the structure of the final bound state, resonances and attenuation 
of the barrier by the nuclear mean field can give an extra energy dependency.

The problem is that the extrapolation is  not usually a straight forward
procedure and some physics has to be taken into account. For example, 
for some systems there is an upturn in the $S$-factor at very low energy
due to a pole at threshold when the photon energy of a 
$A(x,E_{\gamma})B$ capture reaction vanishes. This is a quite common
feature and a straight line extrapolation of $S(E)$ from the high energy 
data region to $S(0)$ would not work. 
According to Jennings {\it et al.} \cite{jen98}, the presence of the 
pole  suggests the  $S$-factor may be parametrized as Laurent series
$S=d_1\times E_{\gamma}^{-1} + d_0 + d_1\times E_{\gamma}$, and 
they propose the following expression to parametrize the $S$-factor:

\begin{equation}
\frac{S(E)}{S(0)}= \frac{a}{E_B +E} + b +c\times E \hspace{0.5cm} or
\hspace{0.5cm} S(E)= \frac{A}{E_B +E} + B +C\times E
\end{equation}

A good test for this assumption is the investigation of the low
energy region of the $^{16}$O(p,$\gamma$)$^{17}$F 
capture reaction. This reaction is an important reaction in the CNO cycle 
of our Sun but  also in  evolutionary phase referred as the asymptotic giant
branch (AGB) of some massive stars \cite{her05}. 
A very good and precise set of data has been obtained for this reaction 
for both  ground-state (5/2$^+$) and first excited state (1/2$^+$) 
transitions of $^{17}$F,  as can be seen in Figure 1.
In the figure the upturn for the reaction leading to the 
$1/2^+$ excited state is very clear.

\begin{figure}
\includegraphics[height=9cm,width=7cm,angle=-90]{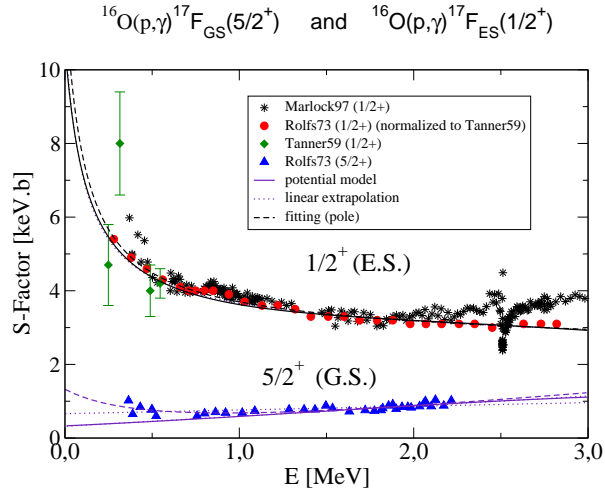}
\caption{$S$-factor for the reaction $^{16}$\textrm{O}%
$(p,\gamma)^{17}\mathrm{F}$. The experimental data are
from Ref. \cite{tanner59,rolfs73,morlock97}.}%
\end{figure}

Let's consider only the data from Rolfs \cite{rolfs73} 
of the transition to the  $^{17}$F$_{GS}$(5/2+) ground-state.
By fitting the data with the expression with pole effect and 
by a straight line we get the following parameters, 
respectively:\\

\noindent {\bf a)} $S(E)= \frac{1.0574}{0.6 +E} -0.4359 +0.4584\times E$, 
giving $S(0)$=1.326 and $\chi^2$=0.194

\noindent {\bf b)} $S(E)=0.66189 + 0.1\times E$, which gives
 $S(0)$=0.66 and $\chi^2$=0.305
\par\hfill

As we can see a very different value for $S(0)$ is obtained depending
on the extrapolation performed. The solution for this problem
is including  more physics in the extrapolation to better determine 
the $S$-factor at low energies.  This can be better done with
another $``fitting''$ method called
R-matrix \cite{descou04}. The idea of the R-Matrix calculations 
is to  parameterize the cross sections with a small number of parameters
and then  used to extrapolate the cross section down to astrophysical 
energies. Since it deals with resonances in the continuum,
it can be considered, to some extent, a generalization of the Breit-Wigner 
formalism. The R-matrix framework 
assumes  that the space is divided into  internal region (with radius match
$a$), where nuclear forces are important, and the external
region, where the interaction between the nuclei is governed
by the Coulomb force only. For more details of this method
we recommend the work of P. Descouvemont in ref. \cite{descou04}.
This method, however,  has some limitations. It doesn't work
for all systems, the radius match $a$ is not clearly determined 
(usually it is arbitrary chosen) and it   is valid only for low level 
density composite nuclei. Moreover,  it has adjusted parameters
which not always has a clear physical meaning.

Other possibility is to use what we call   $``non-Fitting''$ methods.
These  other methods are related to theoretical calculations 
and they determine the cross sections 
from wave functions with, in principle, no data is required. 
Among these methods we have the microscopic cluster 
models and Microscopic $``ab~initio''$ models. In the first, 
a many-body Hamiltonian with embedded cluster are involved in a more complex
calculations. Internal structure has to be taken into account 
and the resonating group method can be applied. In the second, 
 even more complicate calculations are involved since
continuum states, which are very difficult to take into account,
have to be considered. Applications of these two calculations
to the cases of $^7$Li(n,$\gamma$)$^8$Li and $^7$Be(p,$\gamma$)$^8$B
capture reactions can be seen in ref. \cite{descou93}.
Another method also considered as $``non-Fitting''$ method is 
the Potential model, where ANC (Asymptotic Normalization 
Coefficient) can be considered as an approximation. This method is quite 
simple to solve numerically since no structure of the colliding nuclei 
has to be taken into account.

\section{The Potential Model}
\par\hfill

 In the potential model, the direct radiative capture (DRC) of  an $s$- 
and/or $d$-wave nucleon (proton or neutron) by a nucleus $A$, proceeding 
via  E1 transition and leaving the compound nucleus $B$ in its ground state, is 
given by:

\begin{equation}
\sigma^{E1}_{A\rightarrow B} (n,\gamma) = {16\pi \over 9\hbar} k^3_{\gamma}
|<\Psi_{scat}|O^{E1}|\Psi_{bound}>|^2,
\end{equation}

\noindent where   $k_{\gamma}=\epsilon_{\gamma}/\hbar c$ is the
wave number corresponding to a $\gamma$-ray energy  $\epsilon_{\gamma}$, 
$O^{E1}$ stands for the electric dipole operator, the initial-state wave 
function $\Psi_{scat}$ is the incoming nucleon wave function in the 
nucleon-nucleus potential and  $\Psi_{bound}$ is wave function 
which describes the single-particle bound state.

Although the potential model works well for many nuclear reactions of interest
in astrophysics, it is often necessary to pursue a more microscopic approach
 to reproduce experimental data. In a microscopic approach,
instead of the single-particle wave functions one often makes use of overlap
integrals, $I_{bound}(\mathbf{r})$, and a many-body wave function for the 
relative motion, $\Psi_{scat}(\mathbf{r})$.  
The effect of many-body will eventually 
disappear at large distances between the nucleon and the nucleus. 
One thus expects that the overlap function asymptotically matches the 
solution of the Schr\"odinger equation, 
with $V=V_{Coul}$ (Pure Coulomb) for protons and $V=0$ for neutrons. 
 This assumption may  be true only for a very 
peripheral capture reaction.
This approximation, when $r\rightarrow\infty$, is called 
ANC (Asymptotic Normalization Coefficient) and: 

\begin{equation}
I_{bound}(r) =ANC\times \frac{W_{-\eta,l_{b}+1/2}(2\kappa r)}{r} \hspace{1cm} 
({\rm for~~protons})
\end{equation}

\noindent where the binding energy of the $A+x$ system is related to $\kappa$ by means
of $E_{B}=\hbar^{2}\kappa^{2}/2m_{nx}$, $W_{p,q}$ is the Whittaker function
and ANC is the asymptotic normalization coefficient.
The $ANC$ can be  obtained from peripheral transfer reactions whose 
amplitudes contain the same overlap function as the amplitude of the  
corresponding capture reaction of interest \cite{gag02}. 

In terms of overlap integral the direct capture
cross sections are obtained from the calculation of

\begin{equation}
 \sigma_{L,J_{b}%
}^{d.c.} \propto|<  I_{bound}(r)||r^{L}Y_{L}|| \Psi_{scat}(r)>|^{2}.
\end{equation}

However, it has been shown that $s$-wave neutron capture, even at rather 
low energies, is not peripheral \cite{men95,nag05} and so it is necessary 
to calculate the wave function 
of the incoming neutron or proton and the wave function for the bound system. 
Thus, in the potential model, it is necessary to calculate the overlap function
also taking into account the internal part of the nuclear potential,
 I$_{bound}$=$SF^{1/2}\times \Psi(r)_{bound}$. 
Here, $SF^{1/2}$ is the spectroscopic amplitude and $\Psi(r)_{bound}$ is 
the wave-function which describes the bound state.

In the potential model, the continuum wave function, $\Psi_{scat}$,  has to be 
calculated with a potential which includes also the nuclear interaction.
Thus, the essential ingredients in the potential model are the potentials 
used to generate the wave functions  $\Psi_{scat}$ and $I_{bound}$, and the 
normalization for the latter which is given by its spectroscopic factor.
Here we use a Woods-Saxon (WS) parameterization to build up the potentials
$V_{0}(r)$ and $V_{SO}(r)$, where for the latter we consider the derivative 
of the  WS form factor.  The
parameters $V_{0}$, $V_{S0}$, $R_{0}$, $a_{0},$ $R_{S0}$, and
$a_{S0}$ are chosen to reproduce the ground state energy $E_{B}$ (or
the energy of an excited state). For this purpose, we define typical
values (Table I) for $V_{S0}$, $R_{0}$, $a_{0},$ $R_{S0}$, and vary
only the depth of the central potential, $V_{0}$. A different set of potential 
depths might be used for continuum states.
 To calculate the non-resonant  part of these capture reactions in the 
framework of the  potential model we used the computer code RADCAP developed 
by Bertulani \cite{ber03}.

We have tested this potential model by determining the  
$^{16}$O(p,$\gamma$)$^{17}$F$_{gs}$, and  $^{16}$O(p,$\gamma$)$^{17}$F$_{1st}$
$S$-factor as a function of energy.
In Table-I we list all the parameters of the 
potentials used to generate the incoming and bound wave functions. 
All the potentials were assumed to be a Woods-Saxon shape with
geometric parameters $r_0=1.25$ and $a=0.65$ fm.
The $J_{B}=5/2^{+}$ ground state ($J_{B}=1/2^{+}$ excited state) of $^{17}$F 
is described as a $j_{B}=d_{5/2}$ proton ($j_{B}=s_{1/2}$ proton) coupled to 
the  $^{16}$O core, which has an intrinsic spin $I_{A}=0^{+}$. 
The gamma-ray transitions are dominated by the $E1$ multipolarity and by 
incoming $p$ waves for both states. The M1 and E2 contributions amount to
less than $0.1\%$ of the dominant E1 contribution, as shown in 
Ref. \cite{rolfs73} where a potential model was also used.
The spin-orbit potential depth $-10.0$ MeV and the spectroscopic factor $1.0$
have been used for both states, although $0.9$ and $1.0$ for 
the ground state and the excited state, respectively, are recommended
by Rolfs in Ref. \cite{rolfs73}. On the other hand, Iliadis {\it et al.} 
\cite{ili08}
recommend values close to unity. 
The continuum states potential depth are set as
the same as that of bound states, since no elastic scattering data
is known for this system. Our results are shown in Figure 2 with solid
line, and as we can see, it reproduces quite well the experimental data. 
The parameterization of the potential model calculations for the 
$^{16}$O(p,$\gamma$)$^{17}$F capture reactions with the 
pole expression is given by: 
\par\hfill

\noindent  for $^{17}$F$_{GS}$(5/2+): $S(E)= \frac{0.0011578}{0.6 +E} + 0.3207 
+0.2734\times E$, giving S(0)=0.319 and 
$\chi^2$=0.0139;

\noindent for $^{17}$F$_{ES}$(1/2+):
$S(E)= \frac{0.8826}{0.11 +E} + 2.9640 -0.1005\times E$, with 
S(0)=10.987 and $\chi^2$=0.2019
\par\hfill

Here also we present the results for the 
$^7$Li(n,$\gamma$)$^8$Li$_{gs,1st}$, $^8$Li(n,$\gamma$)$^9$Li$_{gs}$,
$^7$Be(p,$\gamma$)$^8$B$_{gs}$ and $^8$B(p,$\gamma$)$^9$C$_{gs}$

For the $^7$Li(n,$\gamma$)$^8$Li$_{gs,1st}$ capture reaction,
the gamma-ray transitions are
dominated by the $E1$ multipolarity and by incoming $s$ waves and
$d$ waves. The $J_{B}=2^{+}$ ground state ($J_{B}=1^{+}$ first
excited state) of $^{8}$Li is described as a $j_{B}=p_{3/2}$
neutron interacting with the $^{7}$Li core, which has an
intrinsic spin $I_{x}=3/2^{-}$. Here we used  $R_0=R_C=R_{S0}=2.391$ fm for 
the radius parameters.
For the continuum state, the potential depth was obtained from 
the analysis of Nagai \cite{nag05} which was determined from the
 experimental scattering lengths, see Table-I for the values.
The spectroscopic factors adopted where $SF(g.s.)=0.98$ \cite{gui07} 
and $SF(1st)=0.48$ \cite{nag05}  for the ground and first
excited states, respectively. The capture to the first excited state
contributes to less than 5\% of the total cross section. 
The results for this calculation are shown in Figure 2.

For the $^8$Li(n,$\gamma$)$^9$Li$_{gs}$ capture reaction,
the scattering potential parameters for both  entrance channel spins, 
$s=5/2^+,3/2^+$,  for the $^8$Li($2^+$)+n system were obtained by keeping the 
same volume integral per nucleon, J$_V$/A, as those for the entrance channel 
spins, $s=2^+,1^+$,  deduced from the scattering potentials 
of the $^7$Li+n  system \cite{nag05}. The results are presented in Figure 4.
Details of the analysis for this neutron capture reactions  
$^7$Li(n,$\gamma$)$^8$Li$_{gs,1st}$ and $^8$Li(n,$\gamma$)$^9$Li$_{gs}$ 
are published in  Ref. \cite{gui07}.

For the $^7$Be(p,$\gamma$)$^8$B$_{gs}$ capture reaction,
the $J_{B}=2^{+}$ ground state of $^{8}$B is described as a
$j_{B}=p_{3/2}$ proton coupled to the $^{7}$Be core, which
has an intrinsic spin $I_{A}=3/2^{-}$. For this system we adopted 
$a=0.52$ fm and $V_{so}=-9.8$ MeV. This
is the same set of values adopted in Ref. \cite{ber03}. 
The gamma-ray transition is dominated by the $E1$ multipolarity and by
incoming $s$ and $d$ waves. Our results are shown in Figure 3. 
To reproduce the  M1 resonance we considered $V_{SCAT}=-38.14$ MeV 
and $SF=0.7$  (dashed-dotted line) also with the other parameters 
according to Table I. 

For the $^8$B(p,$\gamma$)$^9$C$_{gs}$ no data is available.
The capture process for this reaction is dominated by E1 transitions
from incoming $s$-waves to bound $p$ states in  the
ground state of $^{9}$C($J_{B}=3/2^{-}$), which is
described as a $j_{B}=p_{3/2}$ proton coupled to the
$^{8}$B core, which has an intrinsic spin $I_{A}=2^{+}$.
The spectroscopic factor has been set to $1.0$ as in Ref.
\cite{Mohr03}, where several spectroscopic factor values are compared.
The results are shown in Figure-4.

\begin{figure}[ptb]
\includegraphics[height=3.0in, width=2.14in,angle=-90 ]{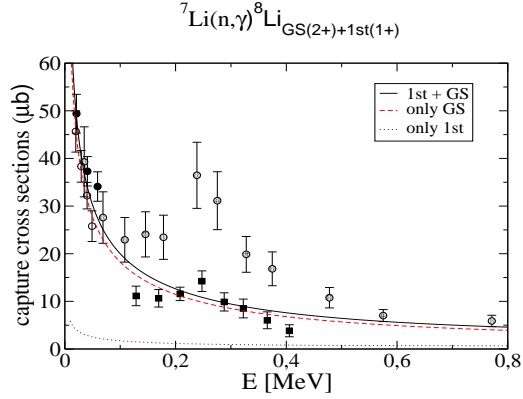}
\caption{Potential model calculations for the reaction 
$^{7}$Li(n,$\gamma$)$^{8}$Li.  Experimental data are from
Refs. \cite{nag05,hei98,nag91,wie892,imh59}.} 
\end{figure}

\begin{figure}[ptb]
\includegraphics[height=2.14 in, width=3.00 in ]{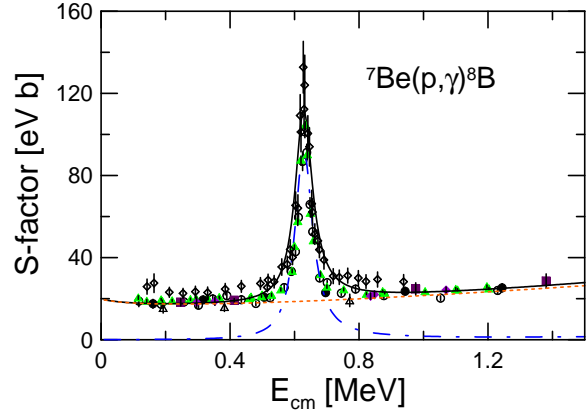}
\caption{(Potential model calculations for the reaction 
$^{7}$Be(p,$\gamma$)$^{8}$B. The dashed-dotted line is the calculation 
for the M1 resonance at $E_{cm}=0.63$ MeV and the dotted line is for the 
non-resonant capture. Experimental data are from
Refs. \cite{VCK70, FED83, Baby03, Junghans03, Iwasa99,Kav69}. 
The total S factor is shown as a solid line.}
\end{figure}

\begin{figure}
\includegraphics[height=2.0in, width=4.5in]{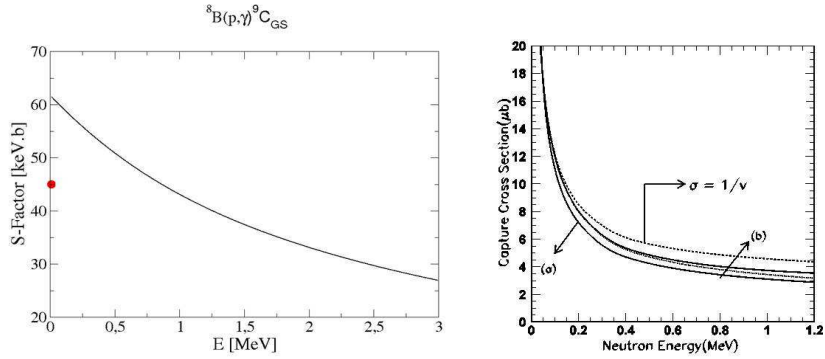}
\caption{(a) Potential model calculations for the 
reaction $^{8}$B(p,$\gamma$)$^{9}$C.  The solid circle symbol at $E=0$ is from  
Refs.  \cite{Trache02}, which were obtained
by ANC (extracted from breakup reaction) calculation.
(b) Potential model calculations for the $^{8}$Li(n,$\gamma$)$^{9}$Li
capture reaction. The lower curves labeled (a) correspond
to the adoption of scattering potential derived from Table-I. 
Curves labeled (b) correspond to the assumption of same
potential for the incoming wave function as for the bound state, for
s-wave neutron only (dotted curve) and s- and d-wave neutrons (solid
curve). The uppermost curve (dashed line) corresponds to a 1/v fit to
the low-energy cross sections.}
\end{figure}

\section{final remarks}
\par\hfill

In this work, we report the results obtained for the non-resonant 
part of the neutron and proton 
capture reactions of light nuclei;  $^7$Li(n,$\gamma$)$^8$Li,
$^8$Li(n,$\gamma$)$^9$Li, $^7$Be(p,$\gamma$)$^8$B  and 
$^8$B(p,$\gamma$)$^9$C in the framework of the potential model.  
As a test of the model we show the results for the 
$^{16}$O(p,$\gamma$)$^{17}$F$_{gs}$(5/2$^+$) and 
$^{16}$O(p,$\gamma$)$^{17}$F$_{ex}$(1/2$+$) capture reactions.
These calculations are part of a more extensive series of calculations
 of neutron and proton capture reactions by light radioactive nuclei
 published elsewhere \cite{huang10}.

Although we have obtained the incoming nucleon scattering 
potentials  for the reaction of interest from analysis of close systems, or 
by considering the same as the bound state, 
it would be interesting to obtain such potentials from  direct elastic 
scattering measurement as $^8$Li+p and $^8$B+p. 
A program of investigation for these elastic
scattering experiments at low energy is under way at the Sao Paulo University
using the radioactive ion beam facility RIBRAS \cite{lic05}.

\begin{table}
\caption{Wood-Saxon potential parameters used in the 
calculations. Depths and B.E. are in MeV 
with $r_0=r_{SO}=r_{coul}=1.25$ fm and $a=a_{SO}=0.65$ fm, where the 
radii are given by $R~=~r_0 \times A_T^{1/3}$. The spin-orbit is given by $V_{SO}$=-10 MeV.}
\begin{tabular}{|cccccc|}
\hline
                       & B.E.  & V$_0$(bound) & SF & channel spin & V$_0$(scat) \\
\hline
$^{16}$O+p=$^{17}$F$_{gs}$(5/2$^+$) & 0.60  & 51.79 & 1.0              &              & 51.79  \\
$^{16}$O+p=$^{17}$F$_{1st}$(1/2$^+$ & 0.11  & 51.42 & 1.0               &              &  51.42 \\
$^7$Be+p=$^8$B$_{gs}$(2$^+$)      & 0.14  & 41.26 & 1.0               &              & 41.26  \\
$^8$B+p=$^9$C$_{gs}$(3/2$^-$)     & 1.30  & 41.97 & 1.0              &               &  41.97 \\
$^7$Li+n=$^8$Li$_{gs}$(2$^+$)     & 2.03  & 46.38 & 0.98(15)        &  $2^+,1^+$     & 56.15,46.50  \\
$^7$Li+n=$^8$Li$_{1st}$(1$^+$)    & 1.05  & 43.30 & 0.48            & $2^+,1^+$     & 56.15,46.50  \\
$^8$Li+n=$^9$Li$_{gs}$(3/2$^-$)   & 4.06  & 47.82 & 0.62(13)      & $5/2^+,3/2^+$ & 58.15,48.15 \\
\hline
\end{tabular}
\end{table}




\bibliographystyle{aipproc}

\end{document}